\documentclass[twocolumn,pre,twoside,showpacs,showkeys,preprintnumbers,floatfix]{revtex4}

\usepackage{amsmath}
\usepackage{amssymb}
\usepackage{graphicx}
\usepackage{dcolumn}
\usepackage{bm}

\def\m#1{\mathrm{#1}}
\def\Ref#1{(\ref{eq:#1})}

\def\p{\partial}
\def\epsilon{\varepsilon}
\def\theta{\vartheta}
\def\rho{\varrho}
\def\erf{\mathop{\mathrm{erf}}}
\def\Int#1#2{\int\!\mathrm{d}^{#1}{#2}\;}
\def\set#1{\underline{#1}}
\def\vec#1{\mathbf{#1}}

\begin{document}

%-------------------------------------------------------------------------------

\title{Relaxation dynamics in fluids of platelike colloidal particles}

\author{Markus Bier}
\email{m.bier@phys.uu.nl}

\author{Ren\'e van Roij}

\affiliation
{
   Institute for Theoretical Physics, 
   Utrecht University, 
   Leuvenlaan 4, 
   3584CE Utrecht, 
   The Netherlands
}

\date{June 8, 2007}

\begin{abstract}
The relaxation dynamics of a model fluid of platelike colloidal particles is investigated by means of a 
phenomenological dynamic density functional theory. The model fluid approximates the particles within the 
Zwanzig model of restricted orientations. The driving force for time-dependence is expressed completely
by gradients of the local chemical potential which in turn is derived from a density functional ---
hydrodynamic interactions are not taken into account. These approximations are expected to lead to qualitatively
reliable results for low densities as those within the isotropic-nematic two-phase region.
The formalism is applied to model an initially spatially homogeneous stable or metastable isotropic fluid which 
is perturbed by switching a two-dimensional array of Gaussian laser beams. Switching on the laser beams leads 
to an accumulation of colloidal particles in the beam centers. If the initial chemical potential and the laser
power are large enough a preferred orientation of particles occurs breaking the symmetry of the laser potential. 
After switching off the laser beams again the system can follow different relaxation paths: 
It either relaxes back to the homogeneous isotropic state or it forms an approximately elliptical high-density 
core which is 
elongated perpendicular to the dominating orientation in order to minimize the surface free energy. For large 
supersaturations of the initial isotropic fluid the high-density cores of neighboring laser beams of the 
two-dimensional array merge into complex superstructures.
\end{abstract}


\maketitle

%========

\section{Introduction}
\label{sec:intro}

Fluids of platelike colloidal particles, e.g., clay suspensions, are of enormous fundamental and technological 
relevance because of their abundance and their distinct properties due to the orientational degrees of freedom 
of the constituting particles. These systems exhibit many interesting phenomena such as flocculation, gelation, 
aging, and even liquid crystal phase transitions \cite{Mourchid1995,Brown1998,Mourchid1998,Bonn1999,Brown1999,%
Levitz2000,Knaebel2000,Abou2001,vanderBeek2003,vanderBeek2004,Wang2005}. During the last decade quite some 
progress has been made in both the synthesis \cite{vanderKooij1998,Liu2003} and the theoretical description of 
the equilibrium structure \cite{Cuesta1999,Rowan2002,Harnau2001,Harnau2002a,Harnau2002b,Bier2004,Harnau2004,%
Costa2005,Harnau2005,Li2005,Bier2005,Bier2006} of suspensions of platelike colloidal particles. However, 
understanding the \emph{non-equilibrium} properties of these systems is still a big challenge. 

Systems out of equilibrium are commonly analyzed either by investigating the \emph{relaxation} of the 
system into equilibrium after an instant change of the conditions, e.g., due to switching an external 
field, or by studying the system in a (non-equilibrium) stationary state. The present theoretical work 
is devoted to the former case of relaxation within a model fluid of platelike colloidal particles; 
stationary states will be addressed in the future. 

In view of the spatial inhomogeneities expected to be found, which in equilibrium systems are most
adequately described by \emph{density functional theory} (DFT) \cite{Evans1979,Evans1989,Evans1991}, 
the current investigation is performed within the framework of a phe\-nom\-eno\-logic\-al \emph{dynamic 
density functional theory} (DDFT), which proposes an equation of motion for the particle number density 
profiles \cite{Dieterich1990}. The latter are assumed to describe the state of the system completely, 
as within DFT any relevant quantity is a functional of the densities. Moreover, DFT is reproduced as the 
stationary limit of DDFT. 

The motivation of the DDFT equation within this work follows the traditional reasoning known from treatments 
of time-dependent Landau-Ginzburg and Cahn-Hillard models in studies of critical dynamics, spinodal decomposition, 
and crystal growth \cite{Langer1971,Kawasaki1977,Collins1985,Harrowell1987}: the larger the thermodynamic 
distance from equilibrium, the faster the change of the state of the system. In the present case the state can 
change due to translation as well as due to rotation of the platelike colloidal particles. Moreover, the total 
number of particles in the system is conserved whereas the orientation is not. Hence the DDFT to be detailed in 
Sec.~\ref{sec:formalism} is analogous to ''model C'' in the classification of Hohenberg and Halperin 
\cite{Hohenberg1977}. Alternatively, the present DDFT can be regarded as an elaborate phase field theory 
\cite{Boettinger2002,Granasy2006} with the order parameter tensor as the natural phase field.

In recent years DDFT equations have been derived based on (overdamped) Langevin dynamics by approximating
the time-dependent two-particle density \cite{Dean1996,Marconi1999,Marconi2000,Archer2004}. As Langevin dynamics
is considered a reasonable description for dilute colloidal suspensions, and as the iso\-tro\-pic-anisotropic
liquid-crystal phase transitions in fluids of highly anisotropic colloidal particles take place at small 
number densities, DDFT is expected to be applicable within the isotropic-nematic two-phase region of fluids of 
platelike colloidal particles.

Here relaxation dynamics is investigated by considering a two-step switching process: First an initially
homogeneous stable or metastable isotropic state is brought out of equilibrium by switching \emph{on} a 
two-dimensional array of 
Gaussian laser beams. The laser potential acts as an external potential which tends to form an inhomogeneous 
equilibrium state. In Sec.~\ref{sec:relaxationinexternalfield} the relaxation towards this inhomogeneous 
equilibrium state is traced by integrating the DDFT equation for the presence of the laser potential. After 
equilibrating the system in the presence of a laser potential, the laser beams are switched \emph{off} and the 
dynamics is described by the DDFT equation for vanishing external field. The system either relaxes back to 
the initial isotropic state or it evolves into an anisotropic state. The various relaxation 
paths will be analyzed in Sec.~\ref{sec:relaxationoffreefluid}. This switching process has been chosen because it 
is expected to be realizable in experiments and, at the same time, numerically convenient boundary conditions can
be used. As an illustration and in order to allow for quantitative
comparison the model parameters have been chosen to describe an aqueous suspension of sterically stabilized 
gibbsite platelets \cite{vanderKooij2000}. 

Section~\ref{sec:discussionandsummary} concludes the present work with a discussion of the applied formalism 
and the numerical results. 

%========

\section{Formalism}
\label{sec:formalism}

%-----------

\subsection{Model fluid}
\label{subsec:model}

The system under consideration is a colloidal suspension of monodisperse hard 
platelike particles dispersed within a continuous solvent. The colloidal 
particles are modeled by square cuboids which can take only one out of 
three mutually perpendicular orientations (Zwanzig approximation 
\cite{Zwanzig1963}), chosen to be parallel to the Cartesian axes. A 
particle is called a $i$-particle if its square face normal 
points along the $i$-axis, $i\in\{x,y,z\}$. The local number density of $i$-particles 
at position $\vec{r}$ is denoted by $\rho_i(\vec{r})$ and the abbreviation 
$\set{\rho}:=(\rho_x,\rho_y,\rho_z)$ is used for convenience. The structure of the model 
fluid is adequately described in terms of the \emph{total density} $\rho:=\sum_i\rho_i$
and the \emph{order parameter tensor} $Q$ \cite{deGennes1993} which, within Zwanzig
models, is given by the diagonal form
\begin{equation}
   Q_{ii'} = \frac{1}{2}\left(3\frac{\rho_i}{\rho}-1\right)\delta_{ii'}.
   \label{eq:Q}
\end{equation}
Here $\delta_{ii'}$ denotes the Kronecker-$\delta$.

%-----------

\subsection{Equilibrium density functional theory}
\label{subsec:edft}

Density functional theory (DFT) \cite{Evans1979,Evans1989} is the method of 
choice in order to investigate equilibrium properties of spatially 
inhomogeneous fluids \cite{Evans1991}. An accurate version of DFT for the Zwanzig
model, which reproduces the exact second and third virial coefficients and which possesses 
the property of dimensional crossover, is the fundamental measure theory due to Cuesta and 
Mart\'{\i}nez-Rat\'{o}n \cite{Cuesta1997.1,Cuesta1997.2}. It has been
applied to investigate monodisperse \cite{Harnau2002b,Martinez1999} and
polydisperse \cite{Martinez2003} Zwanzig models. Within this framework,
the free energy functional is given by
\begin{eqnarray}
   \beta F[\set{\rho}] 
   &\!\!=\!\!&
   \Int{3}{r}\bigg(
   \sum_i\rho_i(\vec{r})\left(\ln(\rho_i(\vec{r})\Lambda^3)-1+\beta V_i(\vec{r})\right) + 
   \nonumber\\
   & &
   \hphantom{\Int{3}{r}\bigg(}
   \Phi(\set{n}(\vec{r}))\bigg),
   \label{eq:freeenergy}
\end{eqnarray}
where $\beta$ is the inverse temperature, $\Lambda$ denotes the thermal de Broglie wavelength, $V_i(\vec{r})$ 
represents the external potential exerted onto $i$-particles at position $\vec{r}$, and
\begin{eqnarray}
   \Phi(\set{n}(\vec{r})) & = & 
   n_0(\vec{r})\ln(1-n_3(\vec{r})) + 
   \nonumber\\
   & &
   \frac{\displaystyle\sum_q n_{1q}(\vec{r})n_{2q}(\vec{r})}{1-n_3(\vec{r})} + 
   \frac{\displaystyle\prod_q n_{2q}(\vec{r})}{(1-n_3(\vec{r}))^2}
\end{eqnarray}
describes the local excess free energy density. The latter is a function of the weighted densities 
$n_\alpha(\vec{r}) := \sum_i \omega_{\alpha,i}\otimes\rho_i(\vec{r})$ with $\otimes$ denoting convolution
and weight functions defined by
\begin{eqnarray}
   \omega_{0,i}(\vec{r})  & = & a(r_x,S_{xi})a(r_y,S_{yi})a(r_z,S_{zi}),
   \nonumber\\
   \omega_{1x,i}(\vec{r}) & = & b(r_x,S_{xi})a(r_y,S_{yi})a(r_z,S_{zi}),  
   \nonumber\\
   \omega_{1y,i}(\vec{r}) & = & a(r_x,S_{xi})b(r_y,S_{yi})a(r_z,S_{zi}),   
   \nonumber\\
   \omega_{1z,i}(\vec{r}) & = & a(r_x,S_{xi})a(r_y,S_{yi})b(r_z,S_{zi}),   
   \nonumber\\
   \omega_{2x,i}(\vec{r}) & = & a(r_x,S_{xi})b(r_y,S_{yi})b(r_z,S_{zi}),   
   \nonumber\\
   \omega_{2y,i}(\vec{r}) & = & b(r_x,S_{xi})a(r_y,S_{yi})b(r_z,S_{zi}),   
   \nonumber\\
   \omega_{2z,i}(\vec{r}) & = & b(r_x,S_{xi})b(r_y,S_{yi})a(r_z,S_{zi}),   
   \nonumber\\
   \omega_{3,i}(\vec{r})  & = & b(r_x,S_{xi})b(r_y,S_{yi})b(r_z,S_{zi}),   
   \label{eq:weightfunctions}
\end{eqnarray}
where the abbreviations $a(r,S):=\frac{1}{2}(\delta(\frac{S}{2}+r)+\delta(\frac{S}{2}-r))$ and
$b(r,S):=\Theta(\frac{S}{2}-|r|)$ are used and $S_{qi}$ denotes the extension of $i$-particles
along the $q$-axis. The term ''fundamental measure theory'' is related to the geometric
interpretation of spatial integrals of the weight functions Eq.~\Ref{weightfunctions}
as particle number, linear extension in $x$-, $y$-, and $z$-direction, cross-sectional area 
perpendicular to the $x$-, $y$-, and $z$-axis, as well as particle volume, respectively.

Within a canonical system the equilibrium density profiles $\set{\rho}^\m{eq}$ minimize 
$F[\set{\rho}]$ under the \emph{constraint} 
\begin{equation}
   \Int{3}{r}\sum_i \rho_i(\vec{r})=\m{const}.
   \label{eq:constraint}
\end{equation} 
The corresponding Lagrange multiplier is the chemical potential $\mu$. 
With the \emph{local chemical potential}
\begin{equation}
   \mu_i\big(\vec{r},[\set{\rho}]\big) := \frac{\delta F}{\delta\rho_i(\vec{r})}\bigg|_{\set{\rho}},
   \label{eq:locchempot}
\end{equation}
$\set{\rho}^\m{eq}$ fulfills the \emph{Euler-Lagrange equation}
\begin{equation}
   \mu_i\big(\vec{r},[\set{\rho}^\m{eq}]\big) = \mu,
   \label{eq:ele}
\end{equation}
i.e., \emph{equilibrium} density profiles render the local chemical potential as a function of 
position ($\vec{r}$) and orientation ($i$) into a constant.

%-----------

\subsection{Dynamic density functional theory}
\label{subsec:ddft}

A system initially out of equilibrium will be driven towards equilibrium. Motivated by 
equilibrium DFT it will be assumed that the state of the system is described by 
\emph{time-dependent} density profiles $\set{\rho}(\cdot,t)$. This assumption implies the
neglect of hydrodynamic interactions, which depend on the velocity field. It is found that
hydrodynamic interactions become more and more relevant for increasing packing fractions 
\cite{Qiu1990,Xue1992}. As the proposed
theory is intended to be applied to dilute colloidal suspensions, taking hydrodynamic interactions
\emph{not} into account is considered to be a reasonable approximation.

In order to investigate
the temporal evolution of the system an \emph{equation of motion} for $\set{\rho}(\cdot,t)$ 
has to be prescribed. Here the following form is proposed:
\begin{equation}
   \frac{\p \rho_i(\vec{r},t)}{\p t} =
   \bigg(\frac{\p \rho_i(\vec{r},t)}{\p t}\bigg)_\m{trans} +
   \bigg(\frac{\p \rho_i(\vec{r},t)}{\p t}\bigg)_\m{rot},
   \label{eq:eom}
\end{equation}
where the first term on the right-hand side describes the contribution due to the translation of
particles with fixed orientation and the second term represents the contribution due to the
rotation of particles keeping the local total density $\rho(\vec{r},t)$ constant.

The translational part fulfills the continuity equation 
\begin{equation}
   \bigg(\frac{\p \rho_i(\vec{r},t)}{\p t}\bigg)_\m{trans} :=
   -\sum_q\frac{\partial j_{iq}(\vec{r},[\set{\rho}(\cdot,t)])}{\partial r_q},
   \label{eq:transcont}
\end{equation}
where $j_{iq}$ describes the translational current of $i$-particles in $q$-direction.
Following Dieterich et al. \cite{Dieterich1990}, the current is assumed to be proportional to the 
local density and to the negative gradient of the local chemical potential:
\begin{equation}
   j_{iq}(\vec{r},[\set{\rho}(\cdot,t)]) :=
   -\Gamma_{iq}\rho_i(\vec{r},t)\frac{\partial\beta\mu_i(\vec{r},[\set{\rho}(\cdot,t)])}{\partial r_q}.
   \label{eq:transcurr}
\end{equation}
Here the \emph{translational diffusion matrix} $\Gamma$ with
\begin{equation}
   \Gamma_{iq} = 
   \left\{
   \begin{array}{ll}
      \Gamma_\|    & , i=q     \\ 
      \Gamma_\perp & , i\not=q
   \end{array} 
   \right.
   \label{eq:transdiffcoeff}
\end{equation}
accounts for different diffusivity of platelike colloidal particles parallel ($\Gamma_\|$) and perpendicular 
($\Gamma_\perp$) to the symmetry axis.

The rotational part of Eq.~\Ref{eom} is modeled by a master equation 
\begin{equation}
   \bigg(\frac{\p \rho_i(\vec{r},t)}{\p t}\bigg)_\m{rot} =
   \bigg(\frac{\p \rho_i(\vec{r},t)}{\p t}\bigg)_\m{rot}^\m{gain} +
   \bigg(\frac{\p \rho_i(\vec{r},t)}{\p t}\bigg)_\m{rot}^\m{loss},
   \label{eq:rot1}
\end{equation}
with the ''gain'' term 
\begin{eqnarray}
   & &
   \bigg(\frac{\p \rho_i(\vec{r},t)}{\p t}\bigg)_\m{rot}^\m{gain} 
   \label{eq:gain}\\
   & \sim &
   \sum_{i'}\Big(
   \beta\mu_{i'}\big(\vec{r},[\set{\rho}(\cdot,t)]\big)-
   \beta\mu_i\big(\vec{r},[\set{\rho}(\cdot,t)]\big)
   \Big)\rho_{i'}(\vec{r},t)
   \nonumber
\end{eqnarray}
and the ''loss'' term
\begin{eqnarray}
   & &
   \bigg(\frac{\p \rho_i(\vec{r},t)}{\p t}\bigg)_\m{rot}^\m{loss}
   \label{eq:loss}\\
   & \sim & 
   -\sum_{i'}\Big(
   \beta\mu_i\big(\vec{r},[\set{\rho}(\cdot,t)]\big)-
   \beta\mu_{i'}\big(\vec{r},[\set{\rho}(\cdot,t)]\big)
   \Big)\rho_i(\vec{r},t).
   \nonumber
\end{eqnarray}
Hence
\begin{eqnarray}
   \bigg(\frac{\p \rho_i(\vec{r},t)}{\p t}\bigg)_\m{rot} 
   & := & 
   -\frac{1}{6\tau}\sum_{i'}\big(\rho_i(\vec{r},t)+\rho_{i'}(\vec{r},t)\big)
   \label{eq:rot2}\\
   & &
   \Big(\beta\mu_i\big(\vec{r},[\set{\rho}(\cdot,t)]\big)-\beta\mu_{i'}\big(\vec{r},[\set{\rho}(\cdot,t)]\big)\Big).
   \nonumber
\end{eqnarray}
The proportionality factors of Eqs.~\Ref{gain} and \Ref{loss} equal the rotational
diffusion coefficient, which in terms of the \emph{rotational relaxation time} $\tau$ is given by $\frac{1}{6\tau}$.

In the limit of infinitely thin platelike colloidal particles Brenner \cite{Brenner1974} has worked out
translational and rotational diffusion coefficients. These expressions in the present notation lead to the rotational 
relaxation time
\begin{equation}
   \tau = \frac{2}{9} \beta\eta D^3,
   \label{eq:rotreltime}
\end{equation}
where $\eta$ is the viscosity of the solvent and $D$ denotes the diameter of the platelike colloidal particles, and to
\begin{equation}
   \Gamma_\|\tau = \frac{D^2}{36}, \quad \Gamma_\perp\tau = \frac{D^2}{24}.
   \label{eq:Gamma}
\end{equation}

Equations \Ref{eom}, \Ref{transcont}, \Ref{transcurr}, and \Ref{rot2} together with definition 
Eq.~\Ref{locchempot} and the density functional Eqs.~\Ref{freeenergy}--\Ref{weightfunctions} completely 
specify the dynamic density functional theory (DDFT) for Zwanzig particles to be investigated within this work.

Integration of the equation of motion Eq.~\Ref{eom} with an initial configuration $\set{\rho}(\cdot,0)$
leads to the time-dependent density profile $\set{\rho}$ which contains all spatial and temporal information. 

The proposed DDFT is consistent with equilibrium DFT as any equilibrium state $\set{\rho}^\m{eq}$ 
fulfills Eq.~\Ref{ele} and therefore does not change under the dynamics represented by Eqs.~\Ref{eom}, 
\Ref{transcont}, \Ref{transcurr}, and \Ref{rot2}.

%-----------

\subsection{External potential}
\label{subsec:externalpotential}

Section~\ref{sec:relaxationinexternalfield} studies the relaxation of the model fluid within an 
external field. In view of conceivable experimental realizations this potential is chosen to model
a two-dimensional array of \emph{Gaussian laser beams}. Following the notation of Ref.~\cite{Gussard1992} 
the \emph{force density} onto an inhomogeneous dielectric fluid is given by 
\begin{equation}
   \vec{f}(\vec{r}) = -\frac{1}{2}\epsilon_0\overline{\vec{E}^2}(\vec{r})\nabla\epsilon(\vec{r})
   \label{eq:forcedensity},
\end{equation}
where $\epsilon_0$ is the permeability of the vacuum, $\overline{\vec{E}^2}$ denotes the temporally averaged 
square of the local electric field, and $\epsilon(\vec{r})$ is the local relative permeability.
A single Gaussian beam of width $w$ and power $P$ propagating parallel to the $z$-axis is described by 
\cite{Gussard1992}
\begin{equation}
   \overline{\vec{E}^2}(\vec{r}) = \frac{2P}{\pi n \epsilon_0 c w^2}\exp\bigg(-\frac{2(r_x^2+r_y^2)}{w^2}\bigg)
   \label{eq:E2}
\end{equation} 
with the refractive index $n$ and the speed of light $c$. The total force $\vec{F}_i(\vec{r})$ exerted on a single 
$i$-particle at position $\vec{r}$ is given by an integration of $\vec{f}$ over the particle volume. Given the 
refractive indices of the solvent $n_\m{solv}$ and of the colloidal particle $n_\m{coll}$, 
$\epsilon$ can be assumed to \emph{discontinuously} change from $n_\m{solv}^2$ outside the particle to 
$n_\m{coll}^2$ inside the particle. The aforementioned volume integration hence reduces to a surface 
integration which can readily be performed in the case of Zwanzig particles. It is found that 
$\vec{F}_i(\vec{r}) = -\nabla V_i(\vec{r})$ with the \emph{laser potential} of a single beam
\begin{equation}
   V_i(\vec{r}) = -\frac{P(n_\m{coll}-n_\m{solv})S_{zi}}{4c}u(r_x,S_{xi})u(r_y,S_{yi}),
   \label{eq:laserpotential}
\end{equation}
where $S_{qi}$ denotes again the extension of $i$-particles along the $q$-axis and 
\begin{equation}
   u(r,S) := \erf\bigg(\frac{\sqrt{2}}{w}\Big(r+\frac{S}{2}\Big)\bigg) -
             \erf\bigg(\frac{\sqrt{2}}{w}\Big(r-\frac{S}{2}\Big)\bigg).
   \label{eq:u}
\end{equation}
For $n_\m{coll}>n_\m{solv}$ the laser potential is attractive.

%-----------

\subsection{Choice of parameters and numerical method}
\label{subsec:choiceofparameters}

In order to obtain solutions of the DDFT equation specified in Sec.~\ref{subsec:ddft} one has to fix the model 
parameters describing the particle geometry, the particle diffusivity, and the laser potential. In the following 
these parameters have been arbitrarily chosen to model the aqueous suspension (viscosity 
$\eta=8.9\cdot 10^{-4}\m{Pa\cdot s}$, refractive index $n_\m{solv}=1.33$) of sterically stabilized gibbsite 
platelets (refractive index $n_\m{coll} = 1.58$) A10P of Ref.~\cite{vanderKooij2000} (diameter $D = 165\m{nm}$, 
aspect ratio $12$) irradiated by a two-dimensional square array of laser beams each of power $P=10\m{mW}$ and 
width $w=5D$ 
propagating in $z$-direction. Neighboring laser beams within the two-dimensional array are chosen to be $30D$ apart. 

Due to the periodicity of the total laser potential the actually infinite system is described in terms of a finite
system with periodic boundary conditions under the influence of one single laser beam centered at the origin.
The density profiles, which vary only in $x$- and $y$-direction but not in $z$-direction, are defined on a 
square of side length $30D$ with periodic boundaries. Due to symmetries of 
the external potential only one quarter has actually to be coded which is done by means of a grid with spacing 
$\frac{D}{2}$. This grid leads to slightly broadened isotropic-nematic interfaces as
compared to calculations with finer grids. However, such a minute inaccuracy is considered irrelevant as compared
to the numerical advantage gained by the relatively coarse grid. A particularly efficient numerical implementation 
is possible if one approximates the colloidal particles by infinitely thin platelets; an exception is 
Eq.~\Ref{laserpotential} where a thickness of $\frac{D}{12}$ is used in order to avoid a vanishing laser potential.

\begin{figure}[!t]
   \includegraphics[scale=0.5]{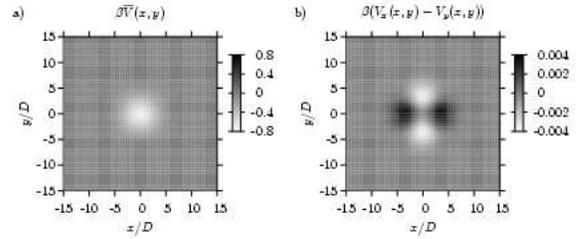}
   \caption{\label{fig:0}a) The orientationally averaged laser potential $\overline{V}(x,y)$ exhibits an approximately
            axial symmetric attraction of platelike colloidal particles by the laser beam. b) Deviations
            from the rotational symmetry appear on a smaller energy scale.} 
\end{figure}
Figure~\ref{fig:0} displays the \emph{orientationally averaged laser potential} $\overline{V} := \frac{1}{3}\sum_iV_i$ 
as well as the local difference $V_x-V_y$ of the potentials exerted onto $x$- and $y$-particles. 
The attractive laser potential is approximately axial symmetric on the energy scale 
$\beta^{-1}$ (Fig.~\ref{fig:0}a) whereas there are deviations on the energy scale $10^{-3}\beta^{-1}$
(Fig.~\ref{fig:0}b).

According to Eq.~\Ref{rotreltime} the rotational relaxation time is given by $\tau\approx 214\m{\mu s}$. The
calculated translational diffusion coefficients (see Eq.~\Ref{Gamma}) compare well with the measured
ones of Ref.~\cite{vanderKooij2000}. The DDFT equation is integrated with respect to time by means of the 
Euler-forward method with integration time steps of $\frac{\tau}{100}$.

For the given set of parameters, one finds a first-order isotropic-nematic bulk phase transition at a reduced 
chemical potential $\mu_\m{b}^* = -1.37271680846513$ with $\mu^* := \beta\mu - 
3\log(\frac{2\Lambda}{D})$. At this binodal (b) the iso\-tro\-pic bulk phase of density 
$\rho_\m{b}^\m{iso}D^3\approx 1.144$ coexists with the nematic bulk phase of density 
$\rho_\m{b}^\m{nem}D^3 \approx 1.576$ and scalar order parameter 
$S = \langle\frac{3}{2}\cos(\theta)^2-\frac{1}{2}\rangle \approx 0.83$, where $\langle\cdot\rangle$ denotes the 
thermal average and $\theta$ is the angle between the particle normal and the nematic director \cite{deGennes1993}.
For illustrative purpose the reduced chemical potentials $\mu^*$ are also expressed in terms of the
\emph{supersaturation} $\sigma := \frac{\rho-\rho_\m{b}^\m{iso}}{\rho_\m{b}^\m{iso}}$ of an isotropic fluid of
reduced chemical potential $\mu^*$.
The isotropic-nematic interfacial tensions for the nematic director pointing parallel and perpendicular to the 
interface normal are given by $\beta\gamma_\|D^2 = 2.051633048\cdot10^{-4}$ and 
$\beta\gamma_\perp D^2 = 3.832557464\cdot10^{-4}$, respectively. The isotropic spinodal (is) is located at the 
reduced chemical potential $\mu_\m{is}^* \approx -1.128$ where the metastable isotropic state takes a density of 
$\rho_\m{is}^\m{iso}D^3 \approx 1.244$. The nematic spinodal (ns) is located at the reduced chemical potential 
$\mu_\m{ns}^* \approx -1.399$ where the stable isotropic state takes a density of 
$\rho_\m{ns}^\m{iso}D^3 \approx 1.133$.

%==========================

\section{Relaxation in external field}
\label{sec:relaxationinexternalfield}

This section describes the influence of the external laser potential $V_j(\vec{r})$ 
(see Sec.~\ref{subsec:externalpotential}) onto an initially homogeneous isotropic fluid of
platelike colloidal particles (see Sec.~\ref{subsec:model}). Depending on the initial chemical potential, two
qualitatively different scenarios are possible.

For reduced chemical potentials $\mu^* < \tilde{\mu}^*$ the fluid stays isotropic throughout and the laser 
potential merely leads to an axial symmetric accumulation of colloidal particles near the origin. For the parameters 
of Sec.~\ref{subsec:choiceofparameters} the limiting reduced chemical potential is $\tilde{\mu}^* \approx -1.6$
(supersaturation $\sigma \approx -0.080$).

\begin{figure*}[!t]
   \includegraphics[scale=0.5]{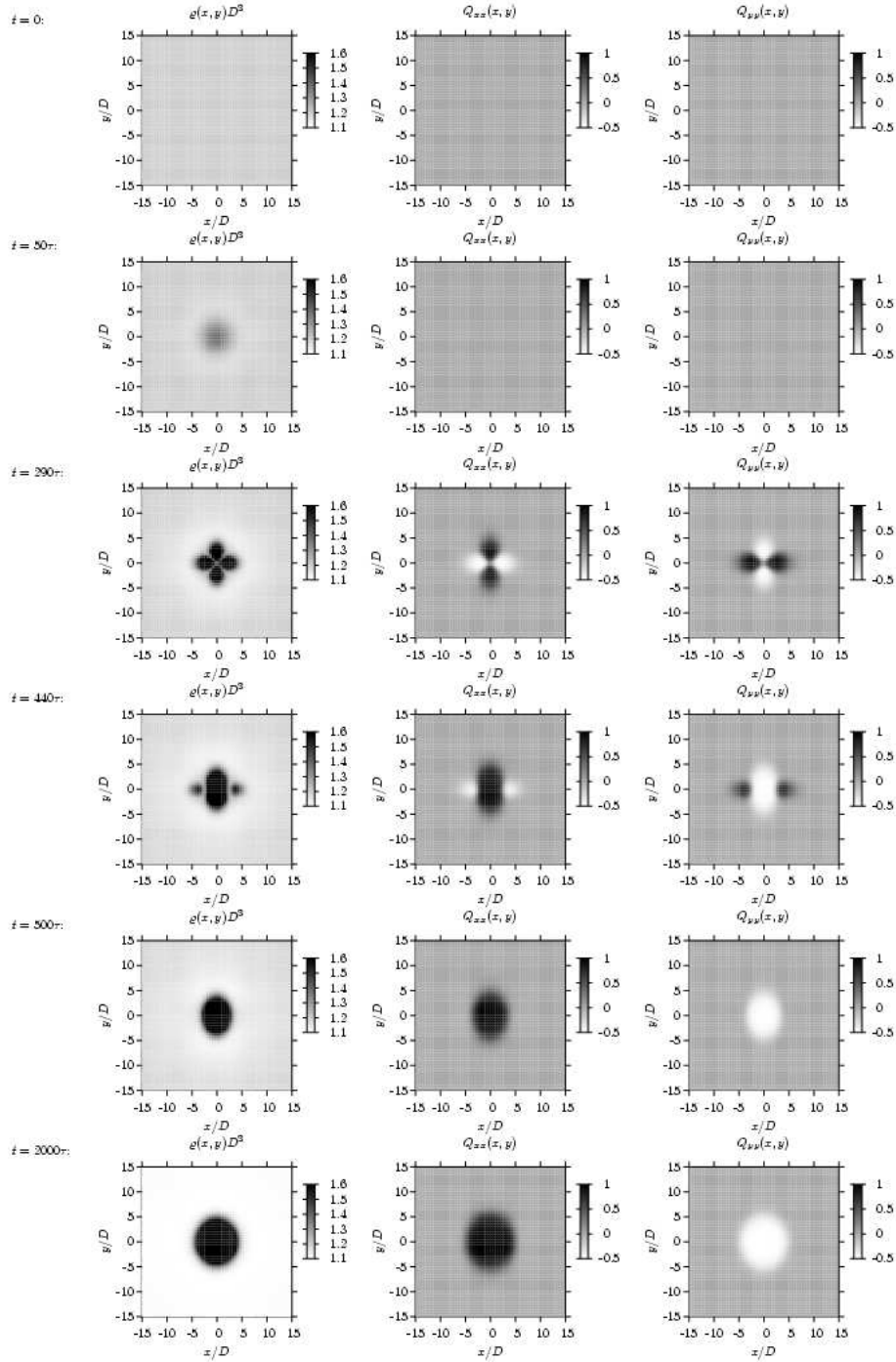}
   \caption{\label{fig:1}Temporal evolution of an initially homogeneous supersaturated isotropic fluid 
            of platelike colloidal particles under the influence of a laser potential in terms of the total density 
            profile $\rho$ and the order parameter tensor component profiles $Q_{xx}$ and $Q_{yy}$. The length 
            scale $D$ and the time scale $\tau$ correspond to the diameter and to the rotational relaxation time 
            of the colloidal particles, respectively. The equilibrium densities at isotropic-nematic two-phase 
            coexistence are given by $\rho_\m{b}^\m{iso}D^3 \approx 1.144$ for the isotropic state and by
            $\rho_\m{b}^\m{nem}D^3 \approx 1.576$ for the nematic state.}
\end{figure*}

However, if the initial reduced chemical potential fulfills $\mu^* > \tilde{\mu}^*$ the initial isotropic 
symmetry of the fluid is broken. Initially homogeneous states within the range $\mu^*\in (\tilde{\mu}^*,\mu^*_{is})$
exhibit the same qualitative behavior. Hence it is sufficient to describe the case of one initial state corresponding
to, e.g., $\mu^* = -1.25$ ($\sigma = 0.044$). A representation of the solutions of the DDFT equation in terms 
of the total density profile $\rho$ and the order parameter tensor component profiles $Q_{xx}$ and $Q_{yy}$ is 
displayed in Fig.~\ref{fig:1}.

After a time of $t=50\tau$ slight distortions of the initial homogeneity due to the attraction of the
colloidal particles by the laser beam become visible in the total density profile $\rho$; the density at the
origin is $\rho(0,0)D^3 \approx 1.382$. Orientational ordering, however, is still negligible.

At time $t=290\tau$ more particles have been concentrated near the origin ($\rho(0,0)D^3 \approx 1.450$) at the 
expense of the direct surrounding. Moreover, four maxima of the total density profile $\rho$ formed near the
origin ($\rho(\pm 1.5D,0)D^3 = \rho(0,\pm 1.5D)D^3 \approx 2.250$). The cubic symmetry of the total density 
profile $\rho$ and the rectangular symmetry of the order parameter components $Q_{xx}$ and $Q_{yy}$ reflect the 
orientation dependence of the external laser potential Eq.~\Ref{laserpotential} (see Fig.~\ref{fig:0}b). 

At $t=440\tau$, the cubic symmetry of the total density profile is broken by forming a bridge between two 
opposite peaks present at time $t=290\tau$. The density at the origin has increased to $\rho(0,0)D^3 \approx 2.723$. 
In the current case this bridge is formed by $x$-particles but a bridge made of $y$-particles is possible as well. 
It has been confirmed that minute rounding-off errors in the very first integration step gives rise to the 
formation of bridges of one or the other orientation. The two peaks of $y$-particles, which do not form a bridge, 
have moved away from the origin and the local maximum of the total density is given by 
$\rho(\pm 3.5D,0)D^3 \approx 1.648$. They are subsequently decreased because the nearby bridge 
forces the particles to take $x$-orientation which afterwards join the bridge.

At time $t=500\tau$ the two peaks of $y$-particles have disappeared completely. The approximately elliptical
core (semiaxes
of half height contour approximately $2.1D$ and $2.6D$) of increased particle concentration near the center 
($\rho(0,0)D^3 \approx 2.805$) comprises mostly $x$-particles. In the following more and more particles from the 
surrounding are added to the high-density core giving rise to a decrease of the total density outside the laser beam. 

From time $t=2000\tau$ onwards the system is almost equilibrated and the state practically does no longer change
in time. The final high-density core has, as the orientationally averaged laser potential $\overline{V}$ 
(Fig.~\ref{fig:1}a), an almost circular cross-section (radius of half height contour approximately $2.9D$).
The total density at the origin is $\rho(0,0)D^3 \approx 3.530$ whereas the minimum of approximately $1.101$ is 
found at a distance of about $8.5D$ from the origin. Due to the small finite size (diameter $30D$) of the 
considered part of the total system, the total density at the (periodic) boundaries has been decreased to a value of 
$\rho|_\partial D^3 \approx 1.104$. 

The trajectory of the system for $\mu^* > \tilde{\mu}^*$ could be summarized as follows: The laser potential,
which is \emph{not} axial symmetric (see Fig.~\ref{fig:0}), attracts particles to the origin and thereby
creates four nematic nuclei (two of $x$-particles and two of $y$-particles) which compete at the origin. If one 
pair of nuclei breaks the symmetry at the origin it forms a bridge whereas the other pair of nuclei decays.
This scenario hinges on the creation of the nematic nuclei and hence on a sufficiently strong laser potential:
The limiting reduced chemical potential $\tilde{\mu}^*$ is expected to decrease upon increasing the laser power $P$.
In particular, for sufficiently strong laser potentials, $\tilde{\mu}^*$ can be located well within the isotropic 
phase of the bulk phase diagram, i.e., the symmetry breaking scenario can occur for initial states that correspond
to a \emph{stable} isotropic fluid. 

For a given initial reduced chemical potential $\mu^*$ one can define a limiting laser power $\tilde{P}(\mu^*)$ 
such that the symmetry breaking scenario will take place if and only if $P > \tilde{P}(\mu^*)$.
For the case $\mu^*=-1.25$ ($\sigma = 0.044$) and the parameters of Sec.~\ref{subsec:choiceofparameters} 
the limiting laser power has been determined approximately as $\tilde{P}(-1.25) \approx 3.5\m{mW}$.

%==========================

\section{Relaxation of the free fluid}
\label{sec:relaxationoffreefluid}

The previous section described the equilibration path of an initially homogeneous isotropic 
fluid within a laser potential. The equilibrium structure in the presence of the laser potential has been
attained after about $t=2000\tau$ (see Fig.~\ref{fig:1}). The present section is concerned with the temporal 
evolution of the fluid after switching off the laser beam at time $t^*:=2000\tau$. Three possible scenarios for 
the evolution of the free fluid have been identified which are determined by the total density of the
initial homogeneous isotropic fluid at time $t=0$.

If the total density at time $t=0$ is small ($\mu^*\lessapprox -1.32,\sigma\lessapprox 0.019$), the high-density 
core present at 
$t=t^*$ (see Sec.~\ref{sec:relaxationinexternalfield}) dissolves completely for $t\rightarrow\infty$ by 
releasing the excess amount of particles to the surrounding. Under these conditions the system relaxes back 
to the initial homogeneous isotropic state. In the case of a supersaturated initial isotropic system 
($\mu^* \in (\mu_\m{b}^*,-1.32), \sigma \in (0,0.019)$, see Sec.~\ref{subsec:choiceofparameters}), classical 
nucleation theory 
would consider the perturbation due to the external potential too small to cross the high free energy barrier 
between the weakly supersaturated metastable isotropic state and a stable nematic state. 

\begin{figure}[!t]
   \includegraphics[scale=0.5]{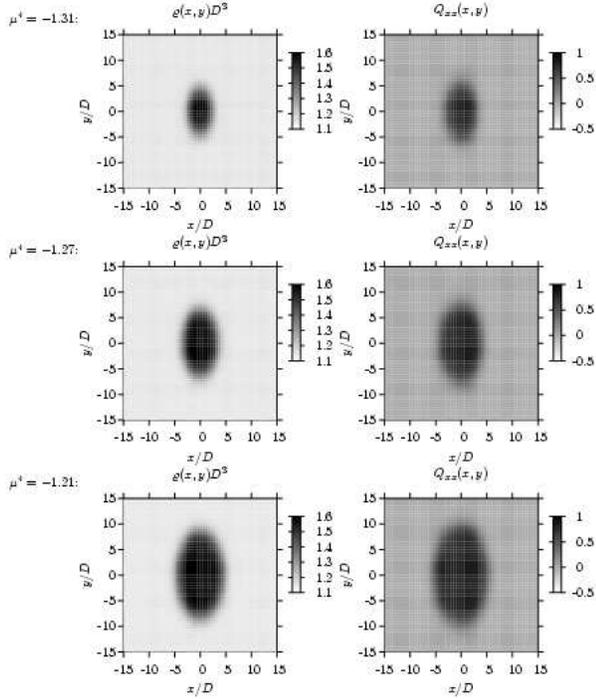}
   \caption{\label{fig:2}Structure of a free fluid of platelike colloidal particles for reduced
            chemical potentials $\mu^*\in\{-1.31,-1.27,-1.21\}$ (supersaturations $\sigma\in\{0.022,0.037,0.058$) at 
            time $t=t^*+10000\tau$. The fluid has been 
            exerted for a time interval $t^*=2000\tau$ to a laser potential (see Fig.~\ref{fig:1}) which afterwards
            has been switched off. These structures represent the equilibrium structures as they do not change
            in time anymore.}
\end{figure}

For intermediate initial supersaturation the high-density core reshapes with time attaining for $t\rightarrow\infty$
a finitely elongated, approximately elliptical shape. Figure~\ref{fig:2} displays the structures of the almost 
equilibrated fluids for initial reduced chemical potentials $\mu^*\in\{-1.31,-1.27,-1.21\}$ (supersaturations 
$\sigma\in\{0.022,0.037,0.058$) at time $t=t^*+10000\tau$. 
These structures exhibit the same total density at the origin $\rho(0,0)D^3 \approx 1.62$, the same total density 
far from the origin $\rho|_\partial D^3 \approx 1.15$, as well as the same aspect ratio $1.85$ of the half height 
contour. Note that the core extension of the equilibrium state can be as small as a few particle diameters (see case 
$\mu^*=-1.31$ ($\sigma=0.022$) in Fig.~\ref{fig:2}). Moreover the high-density core imposes orientational order to 
such a degree
that a preferred alignment of the platelike particles is found even within the dilution zone surrounding it.

An estimate of the aspect ratio of the equilibrium half height contour can be obtained within a \emph{sharp-kink
approximation} which describes the equilibrium structures of Fig.~\ref{fig:2} by a nematic ellipse surrounded
by isotropic fluid where the isotropic-nematic interface is described by a step function.
Assuming an angle-dependent interfacial tension $\gamma(\theta) := \gamma_\|\cos(\theta)^2 + 
\gamma_\perp\sin(\theta)^2$, where $\theta$ denotes the angle between the contour normal and the nematic director,
and where $\gamma_\|$ and $\gamma_\perp$ are the isotropic-nematic interfacial tensions given in
Sec.~\ref{subsec:choiceofparameters}, one obtains an approximate surface free energy $\tilde{F}_\m{surf}$
by integration of $\gamma$ along the contour of an ellipse. Minimizing $\tilde{F}_\m{surf}$ with respect to the 
semiaxes of the ellipse under the constraint of constant area of the ellipse leads to the aspect ratio $1.84249$. 
This value agrees well with the value $1.85$ of the equilibrium structures of Fig.~\ref{fig:2}. The calculation 
also shows that the aspect ratio depends only on the ratio of the interfacial tensions but not on the area of 
the ellipse. 

Within the terminology of classical nucleation theory the perturbation due to the laser potential has been large
enough to cross the free energy barrier between the metastable isotropic state and a state containing an
approximately elliptical nematic core surrounded by an isotropic fluid. In compliance with classical nucleation 
theory, the
shape of the core is determined by a minimum of the surface free energy and the densities within and outside the
core are approximately given by the isotropic-nematic coexistence values. However, whereas classical nucleation theory
and its generalizations \cite{Oxtoby1988} seeks for the fluctuation induced critical nucleus, which corresponds to 
the free energy ''saddle point'' between the metastable and the stable state, the perturbations considered in the 
present work are of external origin and the free energy barrier is crossed along ''higher'' paths.   

\begin{figure}[!t]
   \includegraphics[scale=0.5]{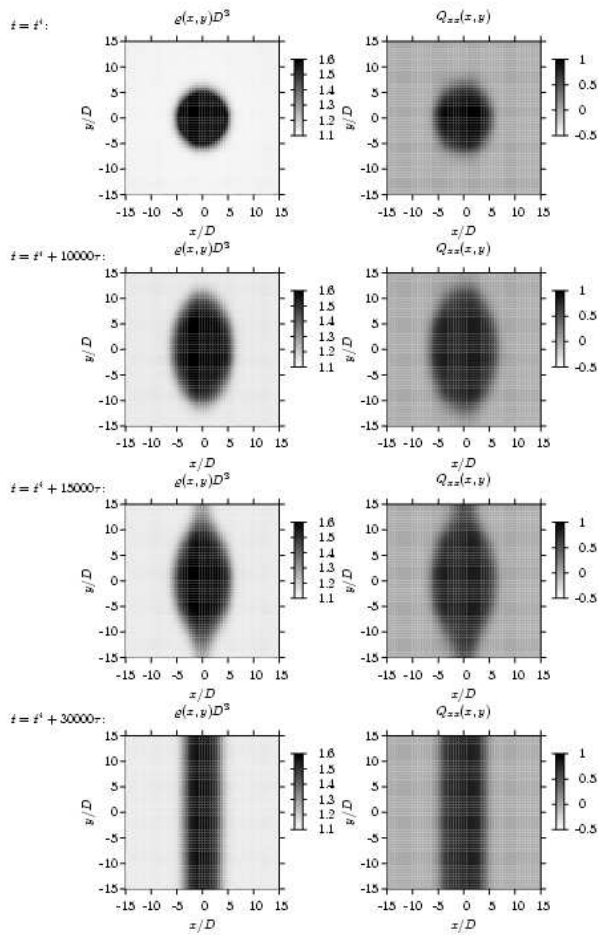}
   \caption{\label{fig:3}Growth of the high-density core within a free fluid of platelike colloidal particles 
            for initial reduced chemical potential $\mu^*=-1.13$ (supersaturation $\sigma=0.087$). The fluid has 
            been exerted for
            a time interval $t^*=2000\tau$ to a laser potential (see Fig.~\ref{fig:1}) which afterwards has 
            been switched off. The growing high-density core ultimately merges with its ''images'' due to periodic 
            boundary conditions.}
\end{figure}

If the initial reduced chemical potential $\mu^*$ is sufficiently large the high-density core contains enough 
particles to grow to the (periodic) boundary of the modeled part of the system where it merges with its 
''images'' into a system-spanning superstructure. Figure~\ref{fig:3} displays this behavior for the case 
$\mu^*=-1.13$ ($\sigma=0.087$).
Directly after switching off the laser beams at time $t=t^*$ a circular high-density core with total density 
$\rho(0,0)D^3 \approx 3.734$ is present. At time $t=t^*+10000\tau$ the total density at the origin has decayed
to $\rho(0,0)D^3 \approx 1.62$ and the aspect ratio of the half height contour is $1.81$. These values agree with
those of the cases displayed in Fig.~\ref{fig:2}. However, the region of orientational order,
which extends beyond the dense core, has reached the periodic boundaries of the system. The high-density core
thereby interacts with its ''images'' which leads to a further stretching as displayed for time $t=t^*+15000\tau$.
Finally the high-density core merges with its ''images'' forming a bandlike structure with a total densities
$\rho(0,0)D^3 \approx 1.58$ and $\rho(\pm 15D,0)D^3 \approx 1.14$. This structure corresponds to a lower free 
energy as compared to the approximately elliptical structure because the isotropic-nematic interface is now 
everywhere 
parallel to the nematic director. The time scale on which the high-density core stretches is much larger than the 
time scales of the processes observed in the presence of the laser potential (see 
Sec.~\ref{sec:relaxationinexternalfield}) because the chemical potential gradients and hence the currents are 
much smaller here. In an experimental realization the high-density cores at the centers of the individual laser 
beams will be oriented in different directions. Upon merging the high-density cores a complex networklike structure
is expected to form.
 
%==========================

\section{Discussion and summary}
\label{sec:discussionandsummary}

The previous three sections introduced and applied a novel dynamic density functional theory (DDFT) for a
model fluid of platelike colloidal particles. The particles are assumed to be infinitely thin and the orientational
degrees of freedom are taken approximately into account by restricting the particle normal to three mutually 
perpendicular directions (Zwanzig model \cite{Zwanzig1963}). Although qualitative, this model exhibits, in
agreement with experimental studies \cite{vanderBeek2003,vanderBeek2004,Wang2005}, an isotropic-nematic bulk 
phase transition at low number densities. Within the framework of DDFT the state is described by a set of 
density profiles and the proposed equation of motion is constructed in order to fulfill the conservation of 
the total number of particles. Time dependence is generated by gradients of the local chemical potential which 
is derived from a density functional. The phenomenological parameters appearing in the equation of motion are 
determined by matching the dilute limit with single particle translational and rotational diffusion 
\cite{Brenner1974}. This approach does not take hydrodynamic interactions between colloidal particles into 
account because they are assumed to be small for the small densities within the isotropic-nematic two-phase 
coexistence region considered here. 

The proposed DDFT is applied to the investigation of relaxation dynamics of the model fluid. In a first step the
relaxation of an initially homogeneous isotropic fluid in an external potential due to a two-dimensional array
of laser beams (Fig.~\ref{fig:0}) is studied. The restriction on the orientational degrees of freedom of the
particles leads to structures of rectangular symmetry (Fig.~\ref{fig:1}). In particular, a transient cubic 
symmetric total density 
profile changes into an only rectangular symmetric total density profile. For a model of continuous orientations 
it can be expected to find an axial symmetry instead of the cubic symmetry. However, the spontaneous symmetry 
breaking into a rectangular symmetric equilibrium state is expected to be found for continuous orientational 
degrees of freedom, too. 

For an infinite system initially prepared within the isotropic-nematic two-phase coexistence region and one 
single laser beam of sufficient strength, instead of an array of laser beams, one would 
expect an equilibrium structure composed of three regions: A narrow high-density core at the position of the laser 
beam is surrounded by an annulus of nematic structure of total density $\rho_\m{b}^\m{nem}$ (see 
Sec.~\ref{subsec:choiceofparameters}) which is surrounded by an annulus of isotropic structure of total density 
$\rho_\m{b}^\m{iso}$ (see Sec.~\ref{subsec:choiceofparameters}); the sizes of the isotropic and nematic annuli 
are related to the initial state by means of the ''lever rule''. In order to form this structure the system has to 
be large enough such that the amount of fluid within the high-density core and the isotropic-nematic interface is 
negligible as compared to the total amount of fluid. However, the two-dimensional grid applied within this work does
not allow for sufficiently large single beam systems. In contrast, the smallness of the periodicity of the system 
considered in this work leads to a rather strong dilution of the fluid in the space between the laser beams.

In classical nucleation theory one considers the temporal evolution of fluctuation induced ''droplets'' of the 
stable (nematic) phase surrounded by the metastable (isotropic) phase, where the densities are chosen as the 
two-phase coexistence values. The smallness of the periodicity of the current system, however, leads to
a high-density core whose size equals the size of the laser beam and whose density is determined by the strength 
of the external potential. Therefore the high-density cores in this work are very different from the ''droplets'' 
considered in classical nucleation theory. Moreover, the relaxation paths here do \emph{not} cross the ''critical
nucleus'', which corresponds to the minimal perturbation that lead to the equilibrium state and which is located 
at a saddle
point of the free energy \cite{Oxtoby1988}. Thus the current results can be partly looked at in terms of the
terminology of classical nucleation theory, because both are based on a free energy function(al),
however, the setup considered here is different from that of classical nucleation theory.

After attaining a state close to the equilibrium state in the presence of the laser potential the laser beams are 
switched off and the relaxation of the (now) free fluid to equilibrium is traced. For sufficiently large initial
supersaturation of the initial state, one finds the central region of increased particle density stretching
(Fig.~\ref{fig:2}). The shape of these cores resemble tactoids which have been found in dispersions of
rodlike colloidal particles \cite{Bernal1941}.
If the moving density fronts hit the periodic boundaries of the modeled part of the system they merge with their 
''images'' into a superstructure (Fig.~\ref{fig:3}). As the nematic directors within the high-density cores
of the individual laser beams may point into different directions the merging of two neighboring cores is 
expected to give rise to one more relaxation process in which the merging cores reorient their colloidal particles
along a common nematic director. This effect, however, is beyond the present study.

In summary, a novel dynamic density functional theory for a model of fluids of platelike colloidal particles
has been proposed and applied to the relaxation dynamics of the fluid under the influence of the potential of
a two-dimensional array of laser beams. A rich phenomenology including the occurrence of individual approximately
elliptical high-density cores and the possible formation of complex superstructures has been found.

%==========================

\begin{acknowledgments}
The authors like to thank Paul van der Schoot for helpful discussions.
This work is part of the research program of the 'Stichting voor 
Fundamenteel Onderzoek der Materie (FOM)', which is financially supported by 
the 'Nederlandse Organisatie voor Wetenschappelijk Onderzoek (NWO)'. 
\end{acknowledgments}

%==========================


\begin{thebibliography}{00}
   \bibitem{Mourchid1995}
      A.\ Mourchid, A.\ Delville, J.\ Lambard, E.\ L\'{e}colier, and P.\ Levitz,
      Langmuir \textbf{11}, 1942 (1995).
   \bibitem{Brown1998}  
      A.\ B.\ D.\ Brown, S.\ M.\ Clarke, and A.\ R.\ Rennie,
      Langmuir \textbf{14}, 3129 (1998).
   \bibitem{Mourchid1998}
      A.\ Mourchid, E.\ L\'{e}colier, H.\ van Damme, and P.\ Levitz, 
      Langmuir \textbf{14}, 4718 (1998).
   \bibitem{Bonn1999} 
      D.\ Bonn, H.\ Kellay, H.\ Tanaka, G.\ Wegdam, and J.\ Meunier,
      Langmuir \textbf{15}, 7534 (1999).
   \bibitem{Brown1999}
      A.\ B.\ D.\ Brown, C.\ Ferrero, T.\ Narayanan, and A.\ R.\ Rennie,
      Eur.\ Phys.\ J.\ B \textbf{11}, 481 (1999).
   \bibitem{Levitz2000}
      P.\ Levitz, E.\ L\'{e}colier, A.\ Mourchid, A.\ Delville, and S.\ Lyonnard,
      Europhys.\ Lett.\ \textbf{49}, 672 (2000).
   \bibitem{Knaebel2000} 
      A.\ Knaebel, M.\ Bellour, M.-P.\ Munch, V.\ Viasnoff, F.\ Lequeux, and J.\ L.\ Harden,
      Europhys.\ Lett.\ \textbf{52}, 73 (2000).
   \bibitem{Abou2001}
      B.\ Abou, D.\ Bonn, and J.\ Meunier,
      Phys.\ Rev.\ E \textbf{64}, 021510 (2001).
   \bibitem{vanderBeek2003}
      D.\ van der Beek and H.\ N.\ W.\ Lekkerkerker,
      Europhys.\ Lett.\ \textbf{61}, 702 (2003).
   \bibitem{vanderBeek2004} 
      D.\ van der Beek and H.\ N.\ W.\ Lekkerkerker,
      Langmuir \textbf{20}, 8582 (2004).
   \bibitem{Wang2005}
      N.\ Wang, S.\ Liu, J.\ Zhang, Z.\ Wu, J.\ Chen, and D.\ Sun, 
      Soft Matter \textbf{1}, 428 (2005).
   \bibitem{vanderKooij1998}
      F.\ M.\ van der Kooij and H.\ N.\ W.\ Lekkerkerker,
      J.\ Phys.\ Chem.\ B \textbf{102}, 7829 (1998).
   \bibitem{Liu2003}
      S.\ Liu, J.\ Zhang, N.\ Wang, W.\ Liu, C.\ Zhang, and D.\ Sun,
      Chem.\ Mater.\ \textbf{15}, 3240 (2003).
   \bibitem{Cuesta1999}
      J.\ A.\ Cuesta and R.\ P.\ Sear,
      Eur.\ Phys.\ J.\ B \textbf{8}, 233 (1999).
   \bibitem{Rowan2002} 
      D.\ G.\ Rowan and J.-P.\ Hansen,
      Langmuir \textbf{18}, 2063 (2002).
   \bibitem{Harnau2001}
      L.\ Harnau, D.\ Costa, and J.-P.\ Harnau,
      Europhys.\ Lett.\ \textbf{53}, 729 (2001).
   \bibitem{Harnau2002a} 
      L.\ Harnau and S.\ Dietrich,
      Phys.\ Rev.\ E \textbf{65}, 021505 (2002).
   \bibitem{Harnau2002b}
      L.\ Harnau, D.\ Rowan, and J.-P.\ Hansen,
      J.\ Chem.\ Phys.\ \textbf{117}, 11359 (2002).
   \bibitem{Bier2004}
      M.\ Bier, L.\ Harnau, and S.\ Dietrich,
      Phys.\ Rev.\ E \textbf{69}, 021506 (2004).
   \bibitem{Harnau2004}
      L.\ Harnau and S.\ Dietrich,
      Phys.\ Rev.\ E \textbf{69}, 051501 (2004).
   \bibitem{Costa2005}
      D.\ Costa, J.-P.\ Hansen, and L.\ Harnau,
      Mol.\ Phys.\ \textbf{103}, 1917 (2005).
   \bibitem{Harnau2005}
      L.\ Harnau and S.\ Dietrich,
      Phys.\ Rev.\ E \textbf{71}, 011504 (2005).
   \bibitem{Li2005}
      L.\ Li, L.\ Harnau, S.\ Rosenfeldt, and M.\ Ballauff, 
      Phys.\ Rev.\ E \textbf{72}, 051504 (2005).
   \bibitem{Bier2005}
      M.\ Bier, L.\ Harnau, and S.\ Dietrich,
      J.\ Chem.\ Phys.\ \textbf{123}, 114906 (2005).
   \bibitem{Bier2006}
      M.\ Bier, L.\ Harnau, and S.\ Dietrich,
      J.\ Chem.\ Phys.\ \textbf{125}, 184704 (2005).
   \bibitem{Evans1979}
      R.\ Evans,
      Adv.\ Phys.\ \textbf{28}, 143 (1979).
   \bibitem{Evans1989}
      R.\ Evans, in \textit{Les Houches, Session XLVIII, 1988 --- Liquides
      aux interfaces / Liquids at interfaces}, edited by J.\ Charvolin, 
      J.\ F.\ Joanny, and J.\ Zinn-Justin (North-Holland, Amsterdam, 1989), 
      p.\ 1.
   \bibitem{Evans1991}
      R.\ Evans, in \textit{Inhomogeneous fluids}, edited by D.\ Henderson
      (Marcel Dekker, New York, 1991), p.\ 89.
   \bibitem{Dieterich1990} 
      W.\ Dieterich, H.\ L.\ Frisch, and A.\ Majhofer,
      Z.\ Phys.\ B \textbf{78}, 317 (1990).
   \bibitem{Langer1971}
      J.\ S.\ Langer,
      Ann.\ Phys.\ \textbf{65}, 53 (1971).
   \bibitem{Kawasaki1977}
      K.\ Kawasaki,
      Prog.\ Theor.\ Phys.\ \textbf{57}, 410 (1977).
   \bibitem{Collins1985}
      J.\ B.\ Collins and H.\ Levine, 
      Phys.\ Rev.\ B \textbf{31}, 6119 (1985) 
      [Erratum: Phys.\ Rev.\ B \textbf{33}, 2020 (1986)].
   \bibitem{Harrowell1987} 
      P.\ R.\ Harrowell and D.\ W.\ Oxtoby,
      J.\ Chem.\ Phys.\ \textbf{86}, 2932 (1987).
   \bibitem{Hohenberg1977}
      P.\ C.\ Hohenberg and B.\ I.\ Halperin,
      Rev.\ Mod.\ Phys.\ \textbf{49}, 435 (1977).
   \bibitem{Boettinger2002}
      W.\ J.\ Boettinger, J.\ A.\ Warren, C.\ Beckermann, and A.\ Karma,
      Annu.\ Rev.\ Mater.\ Res.\ \textbf{32}, 163 (2002).
   \bibitem{Granasy2006}
      L.\ Gr\'{a}n\'{a}sy, T.\ Pusztai, and T.\ B\"{o}rzs\"{o}nyi, 
      in \textit{Handbook of Theoretical and Computational Nanotechnology, Vol.\ 9},
      edited by M.\ Rieth and W.\ Schommers (American Scientific Publishers, Stevenson Ranch, 2006),
      p.\ 525.
   \bibitem{Dean1996}
      D.\ S.\ Dean,
      J.\ Phys.\ A \textbf{29}, L613 (1996).
   \bibitem{Marconi1999}
      U.\ M.\ B.\ Marconi and P.\ Tarazona,
      J.\ Chem.\ Phys.\ \textbf{110}, 8032 (1999).
   \bibitem{Marconi2000}
      U.\ M.\ B.\ Marconi and P.\ Tarazona,
      J.\ Phys.: Condens.\ Matter \textbf{12}, A413 (2000).
   \bibitem{Archer2004}
      A.\ J.\ Archer and R.\ Evans,
      J.\ Chem.\ Phys.\ \textbf{121}, 4246 (2004).
   \bibitem{vanderKooij2000}
      F.\ M.\ van der Kooij, A.\ P.\ Philipse, and J.\ K.\ G.\ Dhont,
      Langmuir \textbf{16}, 5317 (2000).
   \bibitem{Zwanzig1963}
      R.\ Zwanzig, 
      J.\ Chem.\ Phys.\ \textbf{39}, 1744 (1963).
   \bibitem{deGennes1993}
      P.\ G.\ de Gennes and J.\ Prost,
      \textit{The Physics of Liquid Crystals} 
      (Oxford University Press, Oxford, 1993).
   \bibitem{Cuesta1997.1}
      J.\ A.\ Cuesta and Y.\ Mart\'{\i}nez-Rat\'{o}n, 
      Phys.\ Rev.\ Lett.\ \textbf{78}, 3681 (1997).
   \bibitem{Cuesta1997.2}
      J.\ A.\ Cuesta and Y.\ Mart\'{\i}nez-Rat\'{o}n, 
      J.\ Chem.\ Phys.\ \textbf{107}, 6379 (1997).
   \bibitem{Martinez1999}
      Y.\ Mart\'{\i}nez-Rat\'{o}n and J.\ A.\ Cuesta,
      J.\ Chem.\ Phys.\ \textbf{111}, 317 (1999).
   \bibitem{Martinez2003}
      Y.\ Mart\'{\i}nez-Rat\'{o}n and J.\ A.\ Cuesta,
      J.\ Chem.\ Phys.\ \textbf{118}, 10164 (2003).
   \bibitem{Qiu1990}
      X.\ Qiu, X.\ L.\ Wu, J.\ Z.\ Xue, D.\ J.\ Pine, D.\ A.\ Weitz, and P.\ M.\ Chaikin,
      Phys.\ Rev.\ Lett.\ \textbf{65}, 516 (1990).
   \bibitem{Xue1992}
      J.-Z.\ Xue, X.-L.\ Wu, D.\ J.\ Pine, and P.\ M.\ Chaikin,
      Phys.\ Rev.\ A \textbf{45}, 989 (1992).
   \bibitem{Brenner1974}
      H.\ Brenner,
      Int.\ J.\ Multiphase Flow \textbf{1}, 195 (1974).
   \bibitem{Gussard1992}
      R.\ Gussard, T.\ Lindmo, and I.\ Brevik, 
      J.\ Opt.\ Soc.\ Am.\ B \textbf{9}, 1922 (1992).
   \bibitem{Oxtoby1988}
      D.\ W.\ Oxtoby and R.\ Evans,
      J.\ Chem.\ Phys.\ \textbf{89}, 7521 (1988).
   \bibitem{Bernal1941} 
      J.\ D.\ Bernal and I.\ Fankuchen,
      J.\ Gen.\ Physiol.\ \textbf{25}, 111 (1941).
\end{thebibliography}
\end{document}